\documentclass[a4,aip,reprint,jcp,floatfix,groupedaddress]{revtex4-1}
\usepackage{graphicx}
\usepackage{amsmath}
\usepackage{amssymb}
\usepackage{longtable}

\def\s{\sigma} \def\e{\varepsilon}
\def\deg{{^\circ}}
\def\Hprime{H^{\prime}}
\def\R{{\bf R}}
\def\L{L} \def\Lp{L^{\prime}}  \def\Lpp{L^{\prime\prime}}
\def\Rp{{\bf R^{\prime}}} 
 
\def\l{\ell} \def\lp{\ell{^\prime}}
\def\lpp{\ell{^{\prime\prime}}}
\def\TBstrx{{\tilde B}}
 \def\lps{(2\lp+1)!!} \def\ls{(2\l+1)!!} 
  \def\lppms{(2\lpp-1)!!}
\def\evec#1#2{c^{#1}_{#2}}
\def\cevec#1#2{\bar{c}^{#1}_{#2}}

\begin{document}

\title{A tight binding model for water}

\author{A. T. Paxton}
\author{J. J. Kohanoff}
\affiliation{Atomistic Simulation Centre, School of Mathematics
  and Physics, Queen's University Belfast, Belfast BT7 1NN, UK}

\begin{abstract}
We demonstrate for the first time a tight binding model for water
incorporating polarizable anions. A novel aspect is that we
adopt a ``ground up'' approach in that properties of the monomer
and dimer only are fitted. Subsequently we make predictions of
the structure and properties of hexamer clusters, ice-XI and
liquid water. A particular feature, missing in current tight
binding and semiempirical hamiltonians, is that we reproduce the
almost two-fold increase in molecular dipole moment as clusters
are built up towards the limit of bulk liquid. We concentrate on
properties of liquid water which are very well rendered in
comparison with experiment and published density functional
calculations. Finally we comment on the question of the
contrasting densities of water and ice which is central to an
understanding of the subtleties of the hydrogen bond.
\end{abstract}

\maketitle

\section{Introduction}
\label{sec_intro}
We are motivated to construct a model of water for use in
atomistic simulation within the framework of the self consistent
polarizable ion tight binding (TB)
theory.\cite{Finnis98,Paxton09b} The intention is that this
should bridge the outstanding gap between the local density
functional theory (DFT) and classical molecular mechanics
models. The risk is that the model will prove too slow
computationally to replace existing, highly effective classical
models while being unable to capture enough of the physics and
chemistry (especially of the hydrogen bond) to provide a useful
imitation of the first principles approach. To address the latter
concern we have adopted a ``ground up'' strategy in which
parameters of the TB model are fitted first to the monomer and
lastly to the dimer, so that all the properties of the solid and
liquid are {\it predictions} of the model. The former concern may
be settled by observing at once that our model is capable of
propagating 500~self consistent molecular dynamics time steps
for a unit cell of 128 water molecules in one CPU~hour over four
threads of a commodity multicore 2.4GHz processor---12~ps per day
using a time step of 1~fs.

The structure of the paper is this. We begin in
section~\ref{sec_int} by describing TB models constructed
intuitively using parameters that are selected with little
adjustment from published TB models, in particular drawing on
Walter Harrison's seminal scheme.\cite{Harrison80} This should
give the reader confidence that the TB approach contains the
essential physics. In section~\ref{sec_fit} we demonstrate a
model that is contructed by tuning the parameters using a complex
genetic algorithm that provides us with the most accurate
rendering of the properties of the monomer, in particular its
polarisability and force constants, and the dimer, in particular
its geometry and the shape of the energy versus length of the
hydrogen bond as deduced from high level quantum chemical
calculations.  We apply this model first to the structure and
energetics of the water hexamer and to ice
(sections~\ref{sec_hexamer} and~\ref{sec_ice}); and in
section~\ref{sec_liquid} we demonstrate its predictive power in
respect of the properties of liquid water, especially the radial
distribution function, dielectric constant and self diffusion
coefficient. We discuss our results in
section~\ref{sec_discussion} in which we also consider the
question of the relative densities of ice and water and we
conclude in section~\ref{sec_conclusions}.

\section{Intuitive models}
\label{sec_int}
The self consistent polarizable ion tight binding theory was
proposed in the first instance to address the atomic and
electronic structure of ceramic oxides whose properties are
dominated by the highly polarizable oxygen
anion.\cite{Finnis98,Fabris01} Later it was applied to problems
involving polarizability in molecular physics.\cite{Paxton09b}
The method is described in detail in a recent
textbook\cite{Finnis03} and lecture article.\cite{Paxton09a} We
should point out a close similarity between our theory and that
of the SCC-DFTB method.\cite{Elstner98} The principal difference
is that we go beyond the point charge approximation and allow the
self consistent development of point charge multipoles; this is
accompanied by associated higher angular momentum components of
the electrostatic potential allowing us to describe crystal field
effects in a self consistent manner. The second difference is in
our approach to the choice of model parameters, which we do
largely by rational intuition rather than direct computation and
fitting to density functional total energies. In
section~\ref{sec_fit} when we demonstrate our ``genetic'' model,
we will emphasize that for the special case of water, it is
better to fit to high level quantum chemistry and experiment than
to DFT.

\subsection{Monomer}
\label{subsec_int_monomer}

\subsubsection{A simple non self consistent model}

One might construct a very simple, non self consistent model for
the monomer in the spirit of Harrison's ``solid state
table''.\cite{Harrison80} In this way we place a 1$s$ orbital on
the hydrogen atoms and 2$s$ and 2$p$ orbitals on the oxygen with
off diagonal hamiltonian matrix elements, or {\it hopping
integrals} being
\begin{equation*}
V_{\l\lp\chi}=\alpha\>\eta_{\l\lp\chi}\frac{\hbar^2}{2m}\>\frac{1}{r^2}
\end{equation*}
in which $\eta_{ss\s}=-1.4$ and $\eta_{sp\s}=1.84$. Here, $r$ is
the bond length and $m$ the electron mass. $\chi$ stands for
either $\s$ or $\pi$ bonding. The factor $\alpha$ is ours, since
we have have found that $\alpha=\genfrac{}{}{}{1.5}{1}{2}$
produces a HOMO--LUMO gap in better agreement with
experiment. Specifically in order to supress intermolecular O--H
interactions and generally to achieve short ranged hopping,
we adopt a modified scaling known as GSP, after its
authors,\cite{Goodwin89} such that
\begin{equation}
\label{eq_GSP}
V(r)=V(r_0)\left(\frac{r_0}{r}\right)^n
               \exp\left[n\left(-\left(\frac{r}{r_c}\right)^{n_c}
                                +\left(\frac{r_0}{r_c}\right)^{n_c}
                   \right)\right]
\end{equation}
The virtue of the GSP scaling is that is has the same value at
$r_0$ (but not slope) as the Harrison form if we use $n=2$ while
it decays exponentially with distance under the control of the
cut off parameters, $r_c$ and $n_c$; $r_0$ is conventionally the
equilibrium bond length, but as with all the parameters this
could depend on the orbital. For this reason there are implicit
$\{\l\lp\chi\}$ subscripts on the parameters in~(\ref{eq_GSP})
that we have suppressed. In this intuitive model we set $n=2$,
$n_c=4$ and $r_c=1.8r_0$; $r_0$ is our target bond length,
1.8094~bohr. The {\it diagonal} matrix elements of the
hamiltonian in the solid state table\cite{Harrison80} are taken
to be Hartree--Fock term values, namely $\e_s=-1$~Ry on the
H--atom, the values for oxygen being $\e_s=-2.142$~Ry and
$\e_p=-1.038$~Ry. We should point out that in this, and indeed
all our subsequent models, the bond angle is dominated by the
$s-p$~splitting on the oxygen atom. This is clear since in the
absence of on-site $s-p$ hybridisation the bond angle would be
exactly~90$\deg$. Then as $\e_s$ is raised relative to $\e_p$ the
bond angle increases accordingly.\cite{Pettifor95} It is notable
that the bond angle is very accurately rendered using the atomic
term values.

\subsubsection{Intuitive point charge model}

\begin{figure}
\caption{\label{fig_models} A cartoon of the water monomer. The
  target bond angle is $\theta=104.26^{\circ}$, and the bond
  length is $r_0=1.809$~bohr. In~(a), the dipole moment arising
  from charge transfer is indicated and it is a simple matter to
  adjust the charge transfer, $\delta$, by a choice of
  Hubbard--$U$ to achieve the target dipole moment, 1.86~D. If we
  admit that the anion is {\it polarizable} (b) then the self
  consistent problem can also be solved. The charge transfer is
  increased by reducing $U$ so that a larger charge transfer
  derived dipole moment exists which is in turn partially
  canceled by the induced dipole, again to achieve the target
  dipole moment. See the text for details of how this is done.}
\begin{center}
\includegraphics[scale=0.5,viewport= 185 450 594 660,clip]{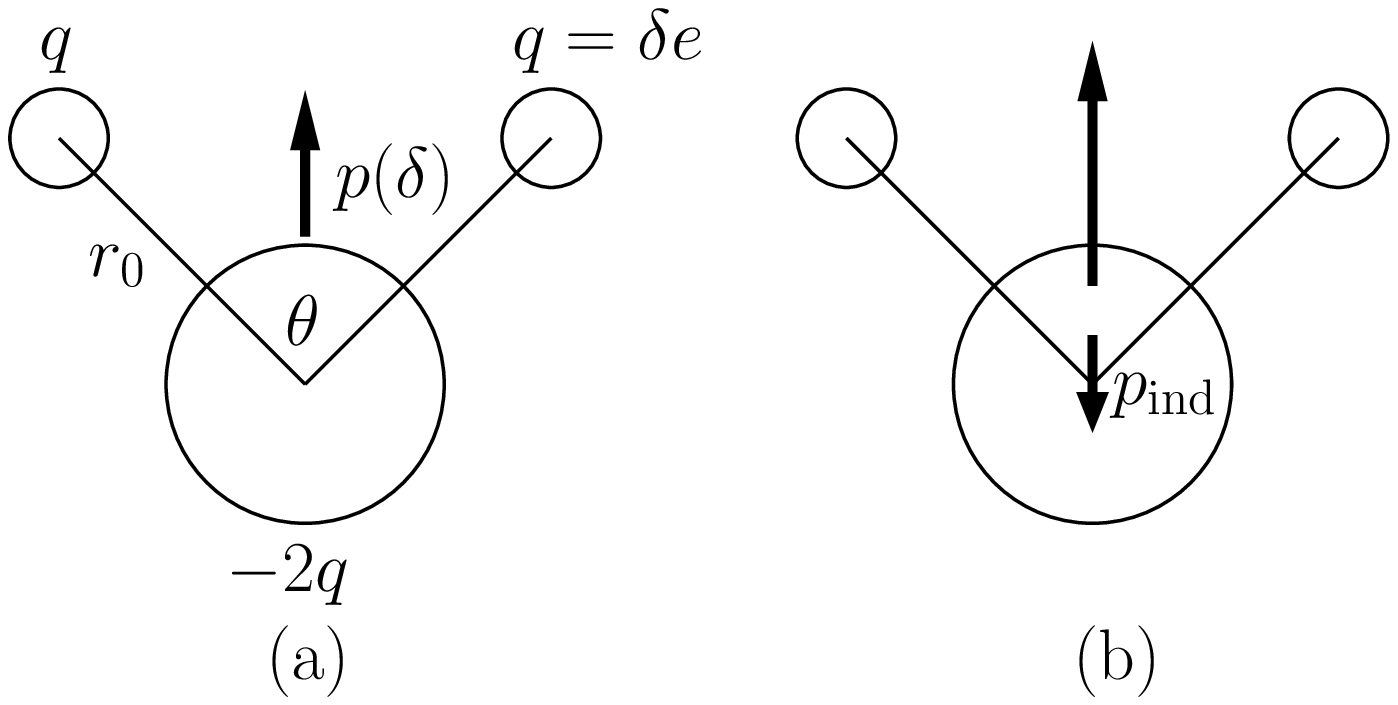}
\end{center}
\end{figure}

We will find that charge transfer which leads to an electrostatic
repulsion between the H-atoms has a very small influence on the
bond angle. Turning to the charge transfer, this is where the
self consistency appears at the simplest level. After solution of
the Schr{\"o}dinger equation using the Harrison TB hamiltonian,
examination of the eigenvectors will reveal Mulliken charges at
the atomic sites. One might ignore these, or compute their
contribution to the energy using elementary electrostatics. Since
charge transfer will always serve to lower the energy, if the
electrostatic potential arising due to Mulliken charges is added
to the hamiltonian and a further solution to the Schr{\"o}dinger
equation is made then, as this process is repeated, charge will
continue to accumulate at the more electronegative site resulting
in a Coulomb catastrophe. This is corrected by a term in the
hamiltonian, an energy proportional to the {\it charge squared},
such that the {\it inter-site} Coulomb energy is balanced by the
on-site {\it Hubbard energy} so called.\cite{Harrison85} These
two are expected largely to cancel in many cases so that it is
admissible to neglect these and stick to a non self consistent
model.\cite{Harrison85} On the other hand the self consistent
inclusion of charge transfer is exactly what is needed to
describe mixed covalent ionic bonding as we expect to find in
water. Therefore at the next level of an intuitive model we
specify Hubbard--$U$ parameters which we choose to be 1~Ry on the
H-atoms and which we adjust on the oxygen until the self
consistent dipole moment has the value, 1.86~Debye~(D), that is
observed. We can now summarise this, our {\it point charge
  model}, by reference to Fig.~\ref{fig_models}(a). A fractional
amount, $\delta$ of an electron is transferred from each H-atom
to the anion; the resulting charges give rise to a dipole moment,
\begin{equation}
\label{eq_pdelta}
p(\delta)=2r_0\delta e\cos\genfrac{}{}{}{1}{1}{2}\theta
\end{equation} 
which, once we have fixed our target bond angle, $\theta$, and
bond length, depends only upon $\delta$ as
$p(\delta)=5.65\delta$~D.  Hence since we know our target dipole
moment it is an easy matter to adjust our one free parameter, the
anion Hubbard--$U$, to obtain a point charge TB model for the
electronic structure. There remains to fix the standard pair
potential, which in the language of the tight binding bond model
is said to account for the all terms except for the bandstructure
energy in the Harris--Foulkes
functional.\cite{Sutton88,Paxton09a} We adopt the GSP
form~(\ref{eq_GSP}) for this allowing two free parameters,
namely, the prefactor which we call $A(r_0)$ and the exponent
$n$, which in the case of a pair potential we call $m$. We fit
these exactly to the bond length and symmetric vibrational force
constant from experiment, while setting $m_c=6$ and
$r_c=2.9$~bohr. This completes the description of our point
charge model.

\subsubsection{Intuitive dipole model}

Let us now admit that the anion is a polarizable species and the
dipole moment created by the charge transfer will result in an
electric field acting at the oxygen atomic site. This will {\it
induce} a dipole moment which we model as a point dipole
\begin{equation}
\label{eq_pind}
p_{\rm ind}=-\alpha_{\rm O}(\delta)\frac{2e\delta}{r_0^2}\cos\genfrac{}{}{}{1}{1}{2}\theta
\end{equation}
in which $\alpha_{\rm O}$ is a notional polarizability of the
oxygen anion. If we knew what this was, and in principle it
should depend on the oxidation state of the anion (as indicated
by its dependence upon $\delta$), then our {\it dipole model}
would follow immediately without any further guessing of
parameters. We would write using~(\ref{eq_pdelta}) and~(\ref{eq_pind})
\begin{equation*}
\frac{p_{\rm ind}}{p(\delta)}=-\frac{\alpha_{\rm O}(\delta)}{r_0^3}
\end{equation*}
while
\begin{equation*}
p(\delta)+p_{\rm ind}=1.86~{\rm D}
\end{equation*}
which could be solved if we knew $\alpha_{\rm O}(\delta)$. In
fact examining the known polarizabilities of the neutral first
row atoms, nitrogen through neon, we found it to be a rather
linear function of the valence leading to an interpolation
formula,
\begin{equation}
\label{eq_interpolate}
\alpha_{\rm O}(\delta)\approx 0.8-0.46\delta\hskip 16pt\hbox{(\AA}^3\hbox{)}
\end{equation}
Unfortunately using the resulting value of $\alpha_{\rm O}$ leads
to a polarization catastrophe: The bond angle tends to zero to
allow the electric field to maximize and induce the largest
possible dipole moment at the anion. Therefore, to construct a
sensible intuitive model, we need to guess a smaller
polarizability and we choose 0.25~$\hbox{\AA}^3$ Thereafter it
should be straight forward to readjust the anion Hubbard--$U$
until the target dipole moment is achieved. However the bond
angle is reduced in this process from the same origin that gives
rise to the polarization catastrophe and it is necessary to raise
the oxygen~$\e_s$. But it is simple enough to adjust these two
parameters simultaneously, after which fitting the pair potential
to $r_0$ and the symmetric vibrational force constant finalizes
the form of our intuitive {\it dipole model}.

\subsubsection{Crystal field}
\label{subsec_crystal-field}
We should make it clear at this point that our self consistent TB
theory does not deal directly with anion polarizability. Instead
the action of the electric field at an atomic site is captured by
a term in the hamiltonian in addition to the non self consistent
hamiltonian such that we have
\begin{equation*}
H = H_0 + \Hprime
\end{equation*}
and $\Hprime$ amounts to a self consistent field having the form
\begin{equation} 
\label{eq_Hprime}
H'_{\R\Lp\,\R\Lpp}=U_{\R}\> q_{\R}\,\delta_{\Lp\Lpp}
               +\sum_{\L}V_{\R\L}\,\Delta_{\lp\lpp\l}\,C_{\Lp\Lpp\L}
\end{equation} 
and the polarizability is expressed through new parameters,
$\Delta_{\lp\lpp\l}$ which are generalizations of the parameters
$\langle r^\l\rangle$ of crystal field
theory.\cite{McClure66,Stoneham75} In~(\ref{eq_Hprime}) $\R$
labels atomic sites and $\L$ is a composite angular momentum
index $\L=\{\l m\}$; $q_{\R}$ is the Mulliken charge transfer at
site $\R$ and $U_{\R}$ is the Hubbard--$U$ at that site. The
second term acts to split the otherwise degenerate atomic energy
levels $\e_{\l}$ through the appearance of on-site, off diagonal
matrix elements of the hamiltonian. $C_{\Lp\Lpp\L}$ are Gaunt
integrals~(\ref{eq_Gaunt}) and $V_{\R\L}$ are expansion
coefficients of the electrostatic potential (energy) seen by an
electron at site $\R$, this potential having been expanded into
spherical waves\cite{Stone96}
\begin{equation*}
V_{\R}({\bf r})=\sum_{\L}V_{\R\L}\>r^{\l}\>Y_{\L}({\bf r})
\end{equation*}
Poisson's equation relates the electrostatic potential to the
multipole moments $Q_{\Rp\Lp}$ of the Mulliken charge transfer
\begin{equation*} 
V_{\R\L}=e^2\sum_{{{\Rp\Lp}\atop{\Rp\neq\R}}}\TBstrx_{\R\L\,\Rp\Lp}\>Q_{\Rp\Lp}
\end{equation*} 
in which $\TBstrx$ is a sort of generalized Madelung matrix,
equal to ${\left\vert\R-\Rp\right\vert}^{-1}$ for point
charges. The general case is described in Appendix~\ref{App_A}.
Finally, the multipole moments are obtained from the self
consistent eigenvectors, $c$, of the hamiltionian\cite{Finnis98,Finnis03,Paxton09a}
\begin{equation*}
Q_{\R\L}=\sum_{\Lp\Lpp}\sum_nf_n\>\cevec{n}{\R\Lp}\evec{n}{\R\Lpp}
        \Delta_{\lp\lpp\l}\>C_{\Lp\Lpp\L}
\end{equation*}
where $f_n$ is the occupation number of state $n$. It is notable
that the crystal field parameter plays a dual role as multipole
strengths in the theory.

For the case of our water model the only polarizability parameter
we need is the quantity $\Delta_{spp}$ at the anion site. The
interpretation of $\Delta_{\lp\lpp\l}$ is that the $\l$-component
of the potential (in the case $\l=1$ the electric field) causes
an on-site coupling in the hamiltonian of the $\lp$ and $\lpp$
orbitals (in this case the $s$ and $p$ are coupled by the crystal
field). In its role as multipole strength parameter, it describes
the contributions of the on-site $\lp$ and $\lpp$ components of
the eigenvectors to the development of the $\l$-pole moment of
the charge (in this case the dipole). This reflects the
well-known requirement to mix $s$ and $p$ orbitals on-site if one
is to develop a dipole moment of the charge.

\subsubsection{Predicted properties of the monomer}
\label{subsec_monomer-predictions}

\begin{table}
\begin{center}
\caption{\label{tbl_monomer-properties} Predictions of the
  intuitive point charge and dipole models for the water
  monomer. The ``genetic model'' is described in
  section~\ref{sec_fit}. Experimental data are taken from the CRC
  Handbook.\cite{CRC02} $E_{\rm gap}$ is the HOMO--LUMO gap.}
\begin{tabular}{lccccccccc}
\hline
 & $\delta$ & $\alpha_{\rm O}$ & $\nu_1$ & $\nu_2$ & $\nu_3$ &$\alpha_{{\rm H}_2{\rm O}}$ & $E_{\rm coh}$ & $E_{\rm gap}$ \\
\noalign{\smallskip}
\hline
\noalign{\smallskip}
model & & $\hbox{(\AA}^3\hbox{)}$ & \multicolumn{3}{c}{force constants (au)} & $\hbox{(\AA}^3\hbox{)}$ & (Ry) & (Ry) \\
\noalign{\smallskip}
\hline
{\bf point}   & 0.33 & ---   & 1.029 & 0.099  & 1.002 & 2.3 & 0.77 & 0.81 \\
{\bf dipole}  & 0.45 & 0.25  & 1.029 & 0.104  & 0.840 & 1.8 & 0.82 & 1.03 \\
{\bf genetic} & 0.47 & 0.27  & 1.029 & 0.065  & 1.061 & 1.5& 0.76 & 0.66 \\
exp.  & ---  & ---  & 1.029 & 0.100  & 1.062 & 1.4 & 0.75 & 0.82 \\
\hline
\end{tabular}
\end{center}
\end{table}

The question now is, what is the predictive power of these two
models? Table~\ref{tbl_monomer-properties} gives an answer. For
each model we show the charge transfer, $\delta$, larger in the
dipole model since its derived dipole moment is to be partially
canceled by the induced moment to achieve the target experimental
value. The symmetric force constant
associated\cite{Footnote1}\nocite{Herzberg45} with the
vibrational frequency $\nu_1$ is fitted in the pair potential,
but the angular and asymmetric frequencies are {\it
  predictions}. First we note that $\nu_2$ is well rendered and
we remind the readers that the angular forces in TB arise from
the $sp\s$ matrix elements, transforming like the Slater--Koster
table.\cite{Slater54} The predictions of the asymmetric force
constant are revealing. It is observed, and also predicted in the
local density functional approximation, that the asymmetric mode
is stiffer than the symmetric. Intuitively one expects the
opposite: since the symmetry is broken there should be greater
opportunity to relax the electronic structure and hence the
curvature of the energy--stretch curve ought to be smaller. In
fact this wrong result is predicted in our models and the effect
is exacerbated in the dipole model for which the direction of
anion polarization is free to rotate and hence lower the
energy. Therefore both models at the intuitive level invert the
ordering of $\nu_1$ and $\nu_3$ and the dipole model is of course
the worse culprit. The total polarizability of the monomer is
well known and can be calculated simply by applying an electric
field and recording the resulting dipole moment.\cite{Paxton09b}
We find that the polarizability is in reasonable agreement with
experiment, the dipole model being quite accurate in this
regard. The cohesive energy of the monomer with respect to its
atoms is rather well reproduced in either model.

\subsection{Dimer}
\label{subsec_int_dimer}
\begin{figure}
\caption{A cartoon of the water dimer. Strictly the two hydrogen
  atoms at the left are eclipsed and $\beta$ is the angle between
  the O--O bond and the plane of the left hand molecule. In the
  instance that $\alpha=0$ we talk of a {\it linear} hydrogen
  bond.}
\begin{center}
\includegraphics[scale=0.5,viewport= 145 530 593 713,clip]{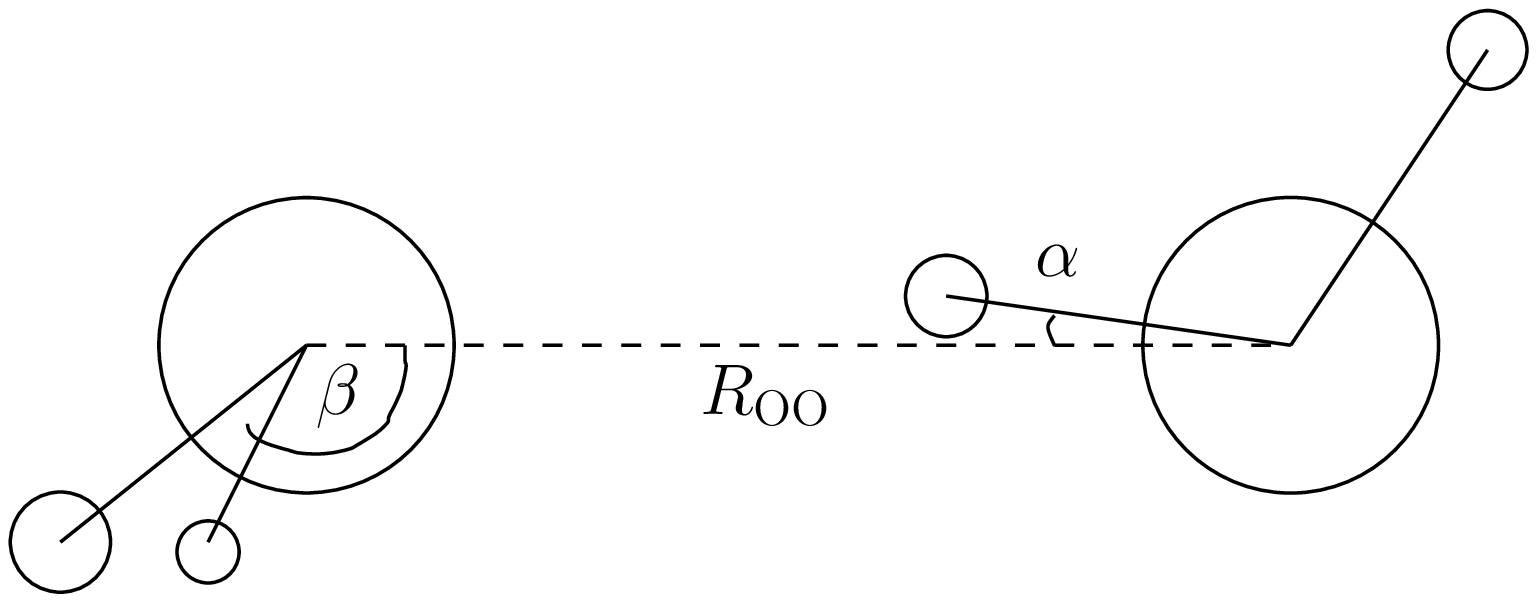}
\end{center}
\label{fig_dimer}
\end{figure}

The curious structure of the water dimer {\it in vacuo} is rather
well known. We show in Fig.~\ref{fig_dimer} a cartoon of the
geometry with the principal length and bond angles indicated. In
order to adapt our intuitive models to describe the dimer we need
to make TB parameters for the O--O interactions. We note that the
O--O distance, marked $R_{\text{OO}}$ in Fig.~\ref{fig_dimer} is
not much different from that in a typical ceramic oxide, such as
zirconia; therefore as a first guess, with some modifications we
try the hopping parameters that we previously used in molecular
statics and dynamics calculations in that
material.\cite{Fabris01,Fabris00} In that work we employed no
O--O pair potential since these anions were second neighbours in
the fluorite lattice. For the intuitive model for water we choose
a simple pair potential of the form
\begin{equation}
\label{eq_expPP}
\varphi(r)=A\,r^{-n}\,e^{-pr}
\end{equation}
having $A$, $n$ and $p$ all greater than or equal to zero. We
adjusted these three parameters by eye to obtain a reasonable fit
of the energy versus O--O bond length (at fixed angles $\beta$
and $\alpha$) calculated with the GAMESS
program\cite{GAMESS1,GAMESS2} using coupled clusters at the
CCSD(T) level. We can then ask, how well do the intutitive models
predict the angles $\beta$ and $\alpha$ shown in
Fig.~\ref{fig_dimer} which have been calculated by Klopper~{\it
  et~al.}\cite{Klopper00} and found to be $\alpha=5.5^{\circ}$
and $\beta=124.4^{\circ}$, while $R_{\text{OO}}=5.50$~bohr. We
use molecular statics to relax the water dimer and we find in the
point charge model, $R_{\text{OO}}=5.34$~bohr,
$\alpha=13.13^{\circ}$ and $\beta=172.7^{\circ}$; while the
dipole model predicts , $R_{\text{OO}}=5.50$~bohr,
$\alpha=5.9^{\circ}$ and $\beta=102.5^{\circ}$ This is very
encouraging. First the models do not predict that the two
molecules line up with their monomer dipoles aligned
antiparallel---a configuration one might well regard as natural
from an elementary electrostatic viewpoint, and one that DFT in
the generalized gradient approximation (GGA) predicts to be
remarkably close (but higher) in energy than the observed
orientation of Fig.~\ref{fig_dimer}. Second, the dipole model at
least makes a good rendering of the angles $\beta$ and $\alpha$
considering the subtlety of the structure. Furthermore the fact
that the dipole model is evidently superior to the point charge
model in this regard argues strongly for the correctness in
accouting for the anion polarizability in the construction of a
TB model for water.

\begin{table}
\begin{center}
\caption{ \label{tbl_parameters} Parameters of our three
  models---intuitive point charge and dipole models and the model
  fitted by genetic evolution. See the text for meanings of the
  various symbols. All quantities are in atomic Rydberg units.}
\begin{tabular}{cccc}
\hline
model:  & {\bf point} & {\bf dipole} & {\bf genetic} \\
\noalign{\smallskip}
\hline
\hline
\multicolumn{4}{c}{On site parameters} \\
\hline
 $\e_s$ (H) & --1  & --1  & --1 \\
 $\e_s$ (O) & --2  & --1.45 & --1.51 \\
 $\e_p$ (O) & --1.038 & --1.038 & --1.20 \\
 $U$ (H)    & 1     & 1   &  1.08 \\
 $U$ (O)    & 0.885 & 0.77 & 1.16 \\
 $\Delta_{spp} $ & --- & --0.84 & --0.924 \\
\hline
\multicolumn{4}{c}{O--H hopping integrals} \\
\hline
 $V_{ss\s}$ & --0.428  & --0.428 &  --0.348 \\
 $V_{sp\s}$ & 0.550   & 0.550    & 0.313 \\
 $n_{ss\s}$ & 2 & 2 & 1.48 \\
 $n_{sp\s}$ & 2 & 2 & 1.98 \\
 $n_c$      & 4 & 4 & 4.04 \\
 $r_c/r_0$  & 1.8 & 1.8 & 1.92 \\
\hline
\multicolumn{4}{c}{O--H pair potential} \\
\hline
  $ A $ &  0.848  &  0.937  &  0.552 \\
  $ m $ & 3.190  & 3.045 &  3.362  \\
  $ m_c $ & 6 & 6 & 6.04 \\
  $ r_c $ & 2.9  & 2.9  & 3.04 \\
\hline
\multicolumn{4}{c}{O--O hopping integrals} \\
\hline
  $V_{ss\s}$ &  --0.072 &  --0.072  &  --0.080 \\
  $V_{sp\s}$ &  0.084   &    0.084  & 0.050 \\
  $V_{pp\s}$ &  0.06    &  0.06     & 0.00012 \\
  $V_{pp\pi}$ & --0.01   &  --0.01  & --0.004 \\
  $n_{ss\s}$ & 2 & 2 & 2 \\
  $n_{sp\s}$ & 2 & 2 & 2 \\
  $n_{pp\s}$ & 3 & 3 & 3 \\
  $n_{pp\pi}$ & 3 & 3 & 3 \\
  $ n_c $     & 6 & 6 & 4 \\
  $ d_0 $     & 5.6  & 5.6  & 5 \\
  $ r_c $     & 9 & 9 & 6.8 \\
\hline
\multicolumn{4}{c}{O--O pair potential} \\
\hline
  $ A $  &  $10^5$  &  $1.5\times10^5$ & --- \\
  $ n $  &  9.7  &  6   & --- \\
  $ p $  &  0 & 1.2 & --- \\
  $ U_1 $ &  --- & --- &  0.010 \\
  $ U_2 $ &  --- & --- &  0.647 \\
  $ r_0 $ &  --- & --- &  5.992 \\
  $ r_1 $ &  --- & --- &  5.494 \\
  $ r_c $ &  --- & --- &  6.110 \\
\hline
\hline
\end{tabular}
\end{center}
\end{table}

\begin{table}
\begin{center}
\caption{\label{tbl_dimer-properties} Properties predicted, or
  fitted, of the water dimer. Target values are taken from
  Klopper~{\it et~al.}\cite{Klopper00}}
\begin{tabular}{ccccc}
\hline
model:  & {\bf point} & {\bf dipole} & {\bf genetic} & target \\
\noalign{\smallskip}
\hline
\hline
 $ R_{\text{OO}} $ (bohr)   & 5.3423 & 5.5011 & 5.5091 & 5.4991 \\
 $\alpha  $ (deg.)   & 13.1   & 5.9    & 3.0    & 5.5    \\
 $\beta   $ (deg.)   & 172.7  & 102.5  & 113.7  & 124.4  \\
 $E_{\rm coh}$ (mRy) & --18.2 & --16.8 & --15.1 & --15.9 \\
 \hline
\hline
\end{tabular}
\end{center}
\end{table}

\begin{figure}
\caption{\label{fig_OO} Cohesive energy verus O--O distance in
  the dimer at fixed $\alpha=5.5^{\circ}$ and
  $\beta=124.4^{\circ}$, using the two intuitive and the fitted
  models, compared to quantum chemical calculations at the
  CCSD(T) level using the GAMESS suite of
  programs.\cite{GAMESS1,GAMESS2}}
\begin{center}
\includegraphics[scale=0.5]{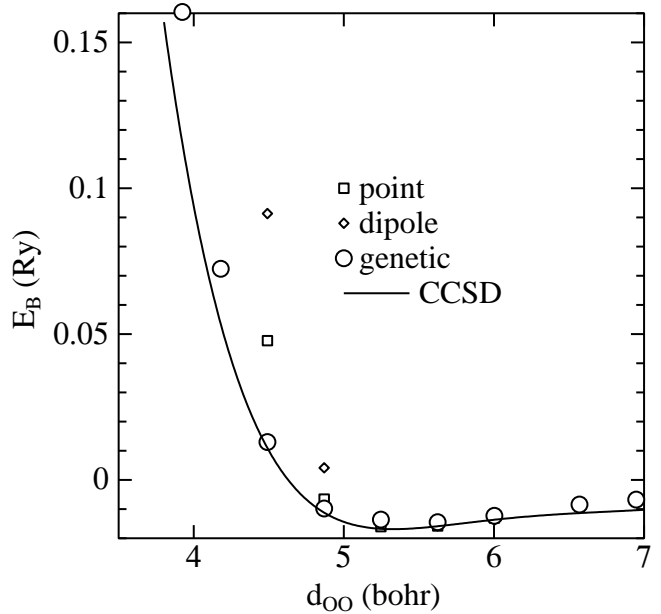}
\end{center}
\end{figure}

\section{Fitted models}
\label{sec_fit}
We hope that the reader is now convinced that the self consistent
polarizable ion TB theory provides a proper framework for the
description of the structure, bonding and energetics of the water
monomer and dimer. In that case there should be no objection to
the next development which is to seek refined hamiltonian matrix
elements, pair potentials and their bond length dependences via a
thorough search in the parameter space. To this end, we have
adopted Schwefel's multimembered evolution
strategy.\cite{Schwefel77,Schwefel93} As we have seen in
section~\ref{sec_int} a set of TB parameters may be used to
predict certain properties. Indeed any set of parameters will
give rise to values of a number of chosen properties which can be
compared to target values taken from experiment or coupled cluster
calculations. A suitably weighted sum of squares of differences
becomes an {\it objective function} and the aim is to find that
set of parameters which gives the smallest value of the objective
function, subject to certain contraints on the parameters so that
they continue to be physically motivated. This last point is very
important---we cannot accept a model whose parameters violate
certain basic truths; for example $\e_p-\e_s>0$, $V_{ss\s}<0$ and
so on. We also have an instinct for the approximate sizes and
appropriate scaling laws for the hopping integrals which are
informed by canonical band theory.\cite{Pettifor95} The procedure
is this. A computer script is written that, for a given set of TB
parameters, calculates the required set of properties and
evaluates the objective function. The evolution algorithm
repeatedly calls this script with TB parameters that belong to
successive parents and offspring until a global minimum is
hopefully found. In the process, a huge set of parameters is
tested and we are free to browse this set to select attractive
looking models and to investigate how certain properties are
governed by particular parameters.

\subsection{Monomer}
\label{subsec_gen_monomer}
In the case of the monomer, the bond length, dipole moment,
frequencies and polarizability are used in the fitting and the
resulting model is added here to
table~\ref{tbl_monomer-properties}. We note that the correct
ordering of $\nu_1$ and $\nu_3$ is acheived, at the
expense of an angular force constant that is significantly too
small. We believe that for the simulations of liquid water, to be
described in section~\ref{sec_liquid} below, the polarizability
of the monomer plays a large role and it is therefore important
at this stage that this is now properly rendered in the ``genetic
model'', so called. 

\subsection{Dimer}
\label{subsec_gen_dimer}
It is a feature of our ``ground up'' philosophy towards the
construction of TB models that, having established a model for
the monomer, we do not further adjust its parameters; and
continue to the fitting of the oxygen--oxygen interactions only
to the properties of the dimer. Our objective function is
constructed using the relaxed geometry of the dimer, in
particular, $R_{\text{OO}}$, $\alpha$ and $\beta$ obtained from a full
geometry optimization. In addition we do calculations of the
cohesive energy at a set of O--O distances at fixed values of
$\alpha$ and $\beta$, namely their target values as found in the
literature.\cite{Klopper00} Five such datapoints are included in
the objective function. Fig.~\ref{fig_OO} shows the energy that
we have calculated\cite{GAMESS1,GAMESS2}
in coupled clusters at the CCSD(T) level compared to data points
from the two intuitive models and our final ``genetic'' model. We
note that the exponential pair potential,~(\ref{eq_expPP}), which
is typical for standard solid state TB models is too steeply
repulsive compared to the CCSD(T) curve. This is clearly a
peculiar signature of the hydrogen bond. We have chosen a weaker
pair potential based on the very first, quadratic, pair potential
proposed by Jim Chadi to describe the harmonic properties of
semiconductor $s-p$ bonding.\cite{Chadi78}
\begin{equation*}
\varphi(r)=U_1\epsilon+U_2\epsilon^2
\end{equation*}
where $\epsilon=(r-r_0)/r_0$ is the fractional change in bond
length relative to some reference distance, $r_0$. Of course a
parabola does not describe the CCSD(T) curve in Fig.~\ref{fig_OO}
at large O--O distances; we replace the potential $\varphi(r)$
with a fifth degree polynomial beyond some distance, $r_1$, which
is constructed to be continuous and twice differentiable at $r_1$
and to vanish smoothly at a cut-off distance, $r_c$. This is
consistent with modern practice in the bond order
potentials.\cite{Znam03} The behavior of this curve at long
distances is dictated by electrostatics and van der Waals
interactions. The latter are explicitly absent in our model while
the electrostatics are included through the self consistent
hamiltonian (section~\ref{subsec_crystal-field}).

All of the parameters of our three models are gathered into
table~\ref{tbl_parameters}. The properties of the dimer are
displayed in table~\ref{tbl_dimer-properties}.

\subsection{The hydrogen bond}
\label{subsec_H-bond}
It is worthwhile to make a few preliminary remarks concerning the
hydrogen bond at this stage of the development. In a sense the
water dimer in its geometry of Fig.~\ref{fig_dimer} is an
archetypal example; and one may ask what is the nature of such a
bond and what determines the values of $R_{\text{OO}}$, $\alpha$ and
$\beta$? The repulsive force which we model with a pair potential
must arise in real life from the closed shell repulsion and
valence--core overlap.  Our view is that the principal attractive
force comes about from the mutual static polarizability of the
two monomers and that in addition the point charge electrostatics
plays a role in the angular disposition of the molecules. On the
other hand we recall again that the O--O distance is only
moderately smaller than that in a typical metal oxide which can
be modeled as mixed ionic--covalent bonding.\cite{Finnis98} For
this reason we include the $sp$ and $pp$ bonding integrals in our
model. If we leave these out, we can make a plausible model, but
we find we cannot reproduce the angles well. One can imagine that
negotiation between the $\s$ and $\pi$ bonds is what results in
the final choice of the angle $\beta$.

\section{The hexamer}
\label{sec_hexamer}

\begin{figure}
\caption{\label{fig_hexamers} Structures of the six hexamer
  isomers considered in the text, relaxed using the TB
  model. The labeling of the oxygen atoms is such that the O--O
  bond lengths can be read off in table~\ref{tbl_hexamer-structure}}
\begin{center}
\includegraphics[scale=0.45]{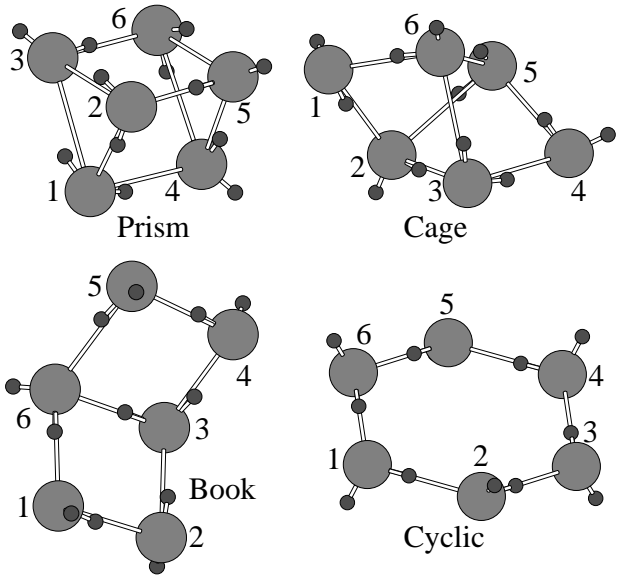}
\end{center}
\end{figure}

The nature of the hydrogen bond as manifested in the dimer can be
further studied in the structure and properties of the hexamer
{\it in vacuo}. Four isomeric structures have been
identified,\cite{Santra08} and remarkably these are practically
degenerate in energy even though their shapes and the number of
hydrogen bonds are very different.\cite{Santra08} The four
isomers are displayed in Fig.~\ref{fig_hexamers} which shows the
structures after geometry optimzation using the genetic fitted TB
model of section~\ref{sec_fit}. As in the dimer, the hydrogen
bonds are not linear: again one sees the angle equivalent to
$\alpha$ in the dimer is some 5$\deg$. The exception to this is
the cyclic hexamer which shows almost exactly linear hydrogen
bonds. Significantly this is the least stable of the
four. Although the TB model overestimates the energy differences
with respect to the most stable, the prism, by as much as an
order of magnitude it is very significant that the ordering
predicted using quantum chemistry at the MP2 level\cite{Santra08}
is reproduced by our model. This is especially noteworthy since
the GGA predicts the ordering quite reversed, namely in that
approximation, the book is most stable followed by the cyclic,
the cage and prism in that order.\cite{Santra08} Our TB model
predicts that the binding energy of the prism hexamer, relative
to the isolated monomers is $-0.132$~Ry. On the other hand the
binding energy relative to three isolated dimers is
$-0.087$~Ry. This reflects the additional binding due first to
the establishment of three more hydrogen bonds; and second, in
our view having greater significance, the attraction due to the
mutual polarization of the water monomers in forming the hexamer
from dimers. In fact we have found, in common with previous first
principles calculations,\cite{DelleSite99,Silvestrelli99} that
the dipole moments of component monomers {\it increase} as they
are assembled into successively larger clusters. This is shown in
table~\ref{tbl_cluster-dipoles}. We will return to this point
when we discuss the structure of liquid water. We should note
that unlike density functional theory, the tight binding model
provides an unambiguous measure of individual molecular multipole
moments; this does not however detract from the fact that these
are, in principle, unmeasurable and therefore ill-defined
quantities.

\begin{table}
\begin{center}
\caption{\label{tbl_cluster-dipoles} Dipole moments on individual
  water molecules found in the monomer, dimer, trimer
  \dots\ hexamer using our fitted genetic, and intuitive point
  charge TB models and compared to available similar results
  using the GGA to density functional
  theory.\cite{Silvestrelli99} Note in connection with remarks in
  section~\ref{sec_liquid} that the point charge model does
  display the expected increase in dipole moment with size of
  cluster, but this is non monotonic, the trimer being an
  exception to the trend, and the magnitude of the effect is
  probably exaggerated by the anomalously large polarizability of
  the monomer (Table~\ref{tbl_monomer-properties}).}
\begin{tabular}{lcccccccc}
\hline
\noalign{\smallskip}
cluster size: & 1 & 2 & 3 & 4 & 5 & 6 & $\cdots$ & liq. \\
\hline
\noalign{\smallskip}
\hline
point   & 1.87 & 2.07 & 2.33 & 2.31 & 2.16 & 2.25 & &     \\
        &      & 2.13 & 2.33 & 2.12 & 2.38 & 2.31 & &     \\
        &      &      & 2.33 & 2.23 & 2.10 & 2.54 & &     \\
        &      &      &      & 2.34 & 2.39 & 2.39 & &     \\
        &      &      &      &      & 2.37 & 2.33 & &     \\
        &      &      &      &      &      & 2.45 & &     \\
mean    & 1.87 & 2.10 & 2.33 & 2.25 & 2.28 & 2.38 & $\cdots$ & 2.51 \\
\hline
genetic & 1.86 & 2.00 & 2.13 & 2.14 & 2.15 & 2.23 & &      \\
        &      & 2.00 & 2.15 & 2.12 & 2.15 & 2.46 & &      \\
        &      &      & 2.14 & 2.26 & 2.31 & 2.55 & &      \\
        &      &      &      & 2.33 & 2.46 & 2.73 & &      \\
        &      &      &      &      & 2.46 & 2.18 & &      \\
        &      &      &      &      &      & 2.50 & &      \\
mean    & 1.86 & 2.07 & 2.14 & 2.21 & 2.31 & 2.44 & $\cdots$ & 3.28 \\
\hline
GGA & 1.87 &  2.1 & 2.4 & --- & ---  & --- &  $\cdots$ & 2.95 \\
\noalign{\smallskip}
\hline
\end{tabular}
\end{center}
\end{table}

The relaxed structures are compared with the MP2 calculations by
displaying the O--O hydrogen bond lengths in
table~\ref{tbl_hexamer-structure}. Our relaxed structures compare
extremely well with the quantum chemistry results; and we remark
that the good agreement and the improvement in our model over the
GGA is because we have rather carefully obtained agreement over a
wide range of O--O bond lengths with the CCSD(T) curve in
Fig.~\ref{fig_OO}. It is well known that the local density
functional theory, even in gradient corrected form, has
difficulty with the energetics of the hydrogen bond and so we
have been able to leap-frog these problems in our TB model by
comparing to the CCSD(T) not the GGA.

\begin{table}
\begin{center}
\caption{\label{tbl_hexamer-structure} Properties of the six hexamer
  isomers considered in the text after geometry optimization
  using the TB model. Energy differences, $\Delta E$, are with
  respect to the Prism isomer. The results labeled ``MP2'' are
  taken from Santra~{\it et al.}\cite{Santra08}}
\begin{tabular}{lcccccccc}
\hline
 &\multicolumn{2}{c}{Prism} & \multicolumn{2}{c}{Cage}
 &\multicolumn{2}{c}{Book} & \multicolumn{2}{c}{Cyclic} \\
 & TB & MP2 & TB & MP2 & TB & MP2 & TB & MP2 \\
\noalign{\smallskip}
\hline
\noalign{\smallskip}
$ \Delta E $ (mRy) & 0 & 0 & 2.8 & 0.2 & 6.9 & 1.5 & 11.8 & 5.0 \\
\noalign{\smallskip}
\hline
\noalign{\smallskip}
 \multicolumn{9}{l}{O--O bond lengths (bohr):} \\
\noalign{\smallskip}
\hline
\noalign{\smallskip}
 1--2 & 4.975 & 5.280 & 5.055 & 5.205 & 4.970 & 5.094 & 5.009 & 5.134 \\
 2--3 & 5.639 & 5.522 & 4.914 & 5.047 & 4.955 & 5.093 & 5.009 & 5.134 \\
 3--1 & 5.597 & 5.545 &  ---  &  ---  &  ---  &  ---  &   --- &  ---  \\
 3--4 &  ---  &  ---  & 5.208 & 5.265 & 5.045 & 5.298 & 5.009 & 5.134 \\
 4--5 & 5.382 & 5.240 & 5.003 & 5.225 & 5.061 & 5.249 & 5.008 & 5.134 \\
 5--6 & 5.564 & 5.424 & 5.542 & 5.520 & 5.299 & 5.261 & 5.009 & 5.134 \\
 6--4 & 5.627 & 5.556 &  ---  &  ---  &  ---  &  ---  &   --- &   --- \\
 1--4 & 5.533 & 5.512 &  ---  &  ---  &  ---  &  ---  &   --- &   --- \\
 2--5 & 4.926 & 5.028 & 5.570 & 5.573 &  ---  &  ---  &   --- &   --- \\
 3--6 & 5.013 & 5.185 & 5.566 & 5.494 & 5.520 & 5.544 &   --- &   --- \\
 6--1 &  ---  &  ---  & 4.998 & 5.149 & 4.972 & 5.081 & 5.009 & 5.134 \\
\noalign{\smallskip}
\hline
\noalign{\smallskip}
\end{tabular}
\end{center}
\end{table}

\section{The structure of ice}
\label{sec_ice}

\begin{figure}
\caption{\label{fig_ice} The structure of ice-XI after geometry
  optimization within our fitted genetic TB model. A unit cell is
marked by straight lines.}
\begin{center}
\includegraphics[scale=0.65,angle=90]{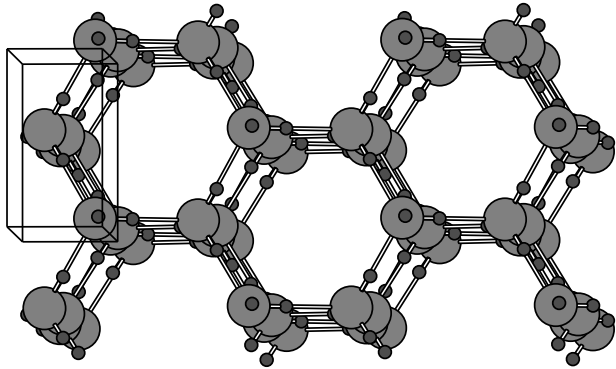}
\end{center}
\end{figure}

We now turn to the solid state to examine how our model
reproduces the structure of ice. This is not the place to make an
exhaustive study of the many solid state phases, nor to attempt a
prediction of the phase diagram, we will leave that to future
work. Most solid phases have disordered hydrogen atom sites, but
the phase Ice-XI is thought to be ``proton-ordered'' so we test
our model against this structure. Hirsch and
Ojam{\"a}e\cite{Hirsch04} have identified a number of putative
structures of Ice~Ih of which Ice-XI is found to have lowest
energy. In Fig.~\ref{fig_ice} we show the structure of Ice-XI
after geometry optimization using our genetic TB model. We find
its cohesive energy with respect to the isolated water monomers
is 39~mRy per monomer, or 12.2~kcal/mol. This result is
consistent with the experimental lattice energy of
13.3~kcal/mol.\cite{Hirsch04} We also find that the optimized
structure of the proton-ordered Ice-Ih ``phase number~6'' of
Hirsch and Ojam{\"a}e\cite{Hirsch04} has an energy higher than
Ice-XI by 0.16~kcal/mol which may be compared with the DMol3/BLYP
calculation result of 0.05~kcal/mol.\cite{Hirsch04} It is
significant that classical force field models find, conversely,
that ``structure number~6'' is more stable than Ice-XI.  Our
optimized Ice-XI has a density at 0$\deg$K of 0.97~g/cm$^3$
compared to the experimental value of 0.93~g/cm$^3$ at
5$\deg$K.\cite{Line96} On the other hand the density of Ice-XI as
calculated in the GGA is as large as 1.00~g/cm$^3$ as we report
in table~\ref{tbl_iceXI-density} which also shows our calculated
lattice constants of ice-XI compared to GGA calculations and
experiment.

\begin{table}
\begin{center}
\caption{\label{tbl_iceXI-density} Lattice constants and density
  of ice-XI at~0$\deg$K calculated in our genetic and intuitive
  point charge TB models compared to GGA
  predictions and experiment. Densities of liquid water are also
  shown for comparison (in g/cm$^3$), and discussed below in
  section~\ref{sec_liquid}. BLYP\cite{Becke88,Lee88} and
  PBE\cite{PBE} denote specific gradient corrected density
  functionals.}
\begin{tabular}{|l|cccc|c|}
\hline
 & \multicolumn{4}{|c|}{Ice-XI} & Liquid \\
 & $a$ & $b$ & $c$ & $\rho$ & $\rho$ \\
\hline
TB (genetic) & 8.422 & 14.506 & 13.668 & 0.967 & 0.926 \\
TB (point)   & 8.488 & 14.754 & 13.900 & 0.928 & 1.030 \\
BLYP         & 8.227\footnotemark[1] & 14.394\footnotemark[1] &
               13.619\footnotemark[1] & 0.995\footnotemark[1] & 0.80\footnotemark[2]\\
PBE          & 8.282\footnotemark[3] & 14.405\footnotemark[3] &
               13.536\footnotemark[3] & 1.000\footnotemark[3] & 0.88\footnotemark[2] \\
experiment   & 8.438\footnotemark[4] & 14.850\footnotemark[4] & 13.780\footnotemark[4] & 0.935\footnotemark[4] & 1.00 \\
\hline
\end{tabular}
\footnotetext[1]{Reference [\onlinecite{Hirsch04}].}
\footnotetext[2]{Reference [\onlinecite{Wang10}].}
\footnotetext[3]{Reference [\onlinecite{Umemoto04}].}
\footnotetext[4]{Measurements at 5$\deg$K.\cite{Line96}}
\end{center}
\end{table}

\begin{figure}
\caption{\label{fig_rdf} Calculated radial distribution functions
  in liquid water, compared to experimental neutron diffraction
  data.\cite{Soper2000} NVE and NPT refer to our genetic dipole
  model simulations in microcanonical and canonical
  ensembles. NVE is done at a density of 1~g/cm$^3$. The average
  density in NPT simulation is displayed in
  table~\ref{tbl_iceXI-density}. We also show NVE simulations in
  our intuitive point charge model. The lowest panel shows the
  integrated quantity $m(r)=\int_0^r g_{\text{OO}}(r')r'^{2}dr'$
  and we mark the position of the first minimum to show the
  average number of O--O bonds. As is generally known this is
  larger than the number, 4, associated with the hexagonal ices,
  such as ice-XI shown in Fig.~\ref{fig_ice}.}
\begin{center}
\includegraphics[scale=0.55]{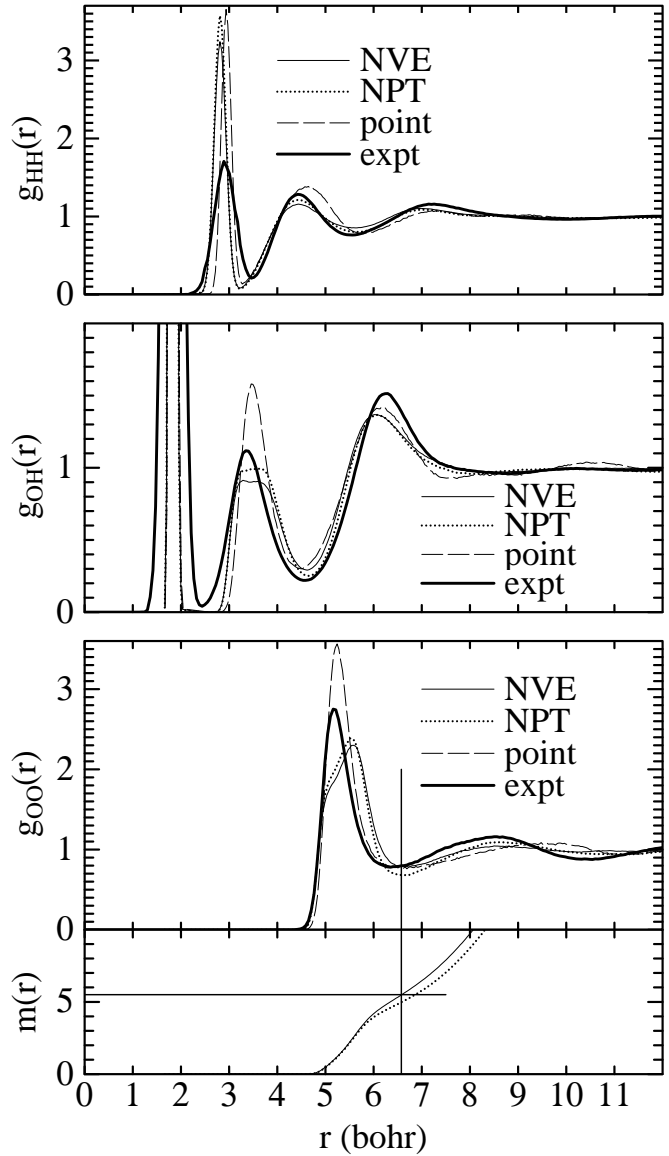}
\end{center}
\end{figure}

\begin{figure}
\caption{\label{fig_dipole-dist} Distribution of the magnitude of
  the dipole monent in an NVE molecular dynamics. In the case of
  the intuitive point charge model there is just the dipole
  moment arising from point charge transfer, indicated by a
  dotted line. In the genetic dipole model the point charge
  contribution is larger, consistent with the picture for the
  monomer, shown in Fig.~\ref{fig_models}; however, unlike in
  the monomer, the partial cancelation from the oppositely
  oriented induced dipole moment is not complete (see
  Fig.~\ref{fig_dipole-cartoon}) so that the total dipole
  moment is on average larger than in the point charge model and
  is consistent with measurement and GGA calculations.}
\begin{center}
\includegraphics[scale=0.4]{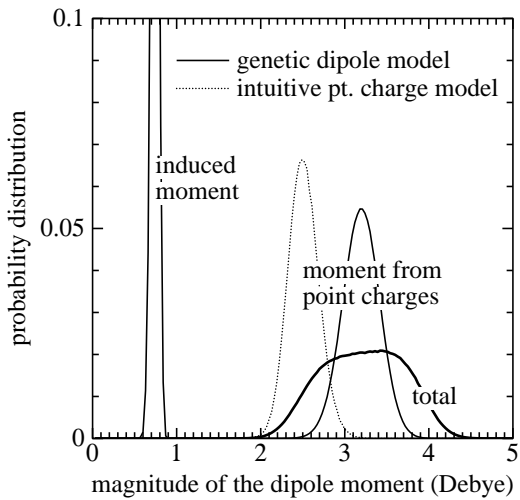}
\end{center}
\end{figure}

\begin{figure}
\caption{\label{fig_dipole-cartoon} A cartoon to illustrate the
  incomplete cancelation of the point charge transfer dipole
  moment in liquid water. The left hand sketch shows molecules in
  the point charge model whose dipoles are necessarily aligned
  along the centerline of the molecule and whose magnitude is
  close to that in the monomer. In the dipole model with induced
  dipoles, the right hand cartoon explains how the molecular
  dipole moment is increased: in the monomer (see
  Fig.~\ref{fig_models}) the induced dipole moment is constrained
  to point opposite to the point charge transfer moment, leading
  to a reduction in the total moment. Conversely in the liquid
  (and also in the solid and clusters discussed above) the
  induced moment ``measures'' the total electric field at the
  anion, not just that due to the two closest hydrogen ions; in
  this way the partial cancelation of the monomer is supressed
  and the resulting total moment is closer to that of the point
  charge transfer which is larger than the corresponding quantity
  in the point charge model.}
\begin{center}
\includegraphics[scale=0.68]{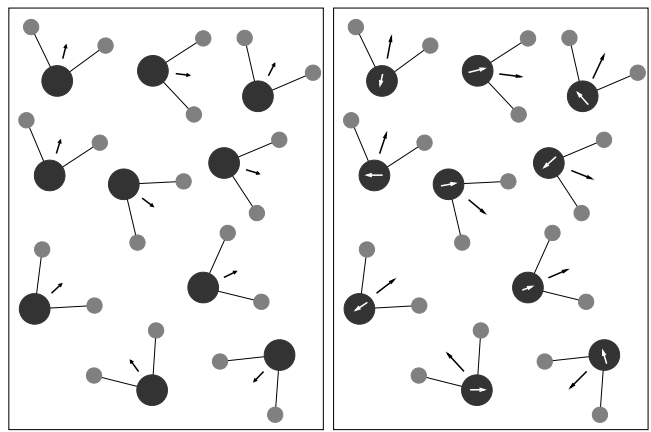}
\end{center}
\end{figure}

\begin{table}
\begin{center}
\caption{\label{tbl_water-properties} Some properties of liquid
  water calculated by our intuitive point charge and genetic
  dipole models, compared to experiment and published GGA
  calculations.  $\bar p_{\text{tot}}$ and $\bar p_{\text{ind}}$
  are total and induced average dipole moments, taken from
  Fig.~\ref{fig_dipole-dist}, $\varepsilon(0)$ is the static
  dielectric constant and $D_{\text{self}}$ is the self diffusion
  coefficient. PBE and BLYP have the meanings of
  table~\ref{tbl_iceXI-density}.}
\begin{tabular}{|l|cc|c|c|}
\hline
       & $\bar p_{\text{tot}}$  & $\bar p_{\text{ind}}$  & $\varepsilon(0)$ & $D_{\text{self}}$ \\
 model & (D) & (D) & & $10^{-5}$ cm$^2$/s \\
\hline
experiment & 2.95$\pm$0.2\footnotemark[1] &  --  & 78,\footnotemark[2] 68\footnotemark[3]   & 2.2\footnotemark[4] \\
genetic (27$^{\deg}$C)   &   3.28        & 0.74 & 86.7 & 3.0 \\
point (27$^{\deg}$C)     &   2.51        &  --  & 58.7 & 3.5 \\
PBE        &   2.95,\footnotemark[5] 3.09\footnotemark[6] &  --  & 67$\pm$6,\footnotemark[6] 75\footnotemark[7]   & 1.6\footnotemark[6] \\
BLYP       & & & & 0.25,\footnotemark[8] 0.55\footnotemark[9] \\
\hline
\end{tabular}
\footnotetext[1]{27$^{\deg}$C, Reference~[\onlinecite{Gubskaya02}].}
\footnotetext[2]{27$^{\deg}$C, Reference~[\onlinecite{Murrell94,Fernandez95}].}
\footnotetext[3]{57$^{\deg}$C, Reference~[\onlinecite{Fernandez95}].}
\footnotetext[4]{27$^{\deg}$C, Reference~[\onlinecite{Eisenberg69}].}
\footnotetext[5]{45$^{\deg}$C, Reference~[\onlinecite{Silvestrelli99}].}
\footnotetext[6]{57$^{\deg}$C, Reference~[\onlinecite{Sharma07}].}
\footnotetext[7]{Extraplolated to 27$^{\deg}$C from data in reference~[\onlinecite{Sharma07}].}
\footnotetext[8]{35$^{\deg}$C, Reference~[\onlinecite{Lee07}].}
\footnotetext[9]{32$^{\deg}$C, Reference~[\onlinecite{Serra04}].}
\end{center}
\end{table}

\section{The structure and properties of liquid water}
\label{sec_liquid}
Finally we apply our TB models to predict properties of liquid
water. For this we take a cubic box of side 29.75~bohr containing
128 water molecules, having periodic boundary conditions and
previously equilibrated in molecular dynamics (MD) using the
SIESTA program and the BLYP~GGA.\cite{Serra04} We further
equilibrated for 100~ps in an NVT ensemble at 27$^{\deg}$C and
1~g/cm$^3$ before an additional 100~ps to gather statistics in an
NVE ensemble. We have also made a simulation in an NPT ensemble
at 27$^{\deg}$C and zero external pressure for a total of 100~ps,
taking statistics from the last 50~ps of simulation time. We
employ the reversible integrators with Liouville operators
developed by Martyna~{\it et al.}\cite{Martyna96} using just a
single Nos\'e--Hoover chain and particle and barostat relaxation
times of 100~fs and 1~ps respectively. We use atomic masses of
oxygen and hydrogen (not deuterium) and time steps of 1~fs in NVT
and NVE and 0.25~fs in NPT ensembles.

First we show radial distribution functions (RDF) in
Fig.~\ref{fig_rdf}. These show the same sorts of features in
comparison to experiment as found in previous calculations using
GGA.\cite{Lee07,Wang10} In particular the first neighbor O--H
peak is narrower than experiment, due probably to the quantum
nature of protons that is usually neglected in MD
simulations. The most critical RDF is that of O--O bonds since
this describes the complex of hydrogen bonds responsible for many
of the subtle properties of water. The RDF from our intuitive
point charge model is very similar to a recent
calculation\cite{Maupin10} using the the SCC-DFTB
method,\cite{Elstner98} which as discussed in
section~\ref{sec_int} above is equivalent in construction to our
point charge model. The point charge model suffers from a nearest
neighbor O--O peak which is too sharp and too high and a lack of
structure in the RDF beyond first neighbors; that is to say,
overstructured in the first solvation shell and understructured
in more distance shells. Both these shortcomings are remedied by
our genetic dipole model at the expense of the first neighbour
O--O peak falling beyond the experiment indicating an average
hydrogen bond length that is too long. It is notable that this
average length is quite insensitive to the simulation box volume,
since the NVE is done at a density of 1~g/cm$^3$, while the
average density in the NPT simulation was 0.926~g/cm$^3$ (see
table~\ref{tbl_iceXI-density}).

We use Figs.~\ref{fig_dipole-dist} and~\ref{fig_dipole-cartoon}
to illustrate the point that we have made earlier (see
Tables~\ref{tbl_cluster-dipoles} and~\ref{tbl_water-properties})
that the molecular dipole moment increases significantly as the
water is built up in clusters from individual molecules having a
dipole moment of 1.86~D to a total dipole moment in liquid water
of about 3~D. It is a particular feature of our dipole model that
this effect is properly captured, in constrast to our point
charge model in which the molecular dipole moments do not grow
monotonically with cluster size (see
Table~\ref{tbl_cluster-dipoles}) and do not reach the expected
$\sim$3~D in the bulk limit. Moreover most of the increase in
dipole moment is probably a result of the polarizability of the
monomer being nearly a factor of two too large in the intuitive
point charge model (see Table~\ref{tbl_monomer-properties}). The
same failing, and to a much greater degree, is apparent in
semiempirical schemes such as AM1 and PM3 (see table~4 of
ref~[\onlinecite{Monard05}]). The quantitative rendering of the
dipole moments as a function of environment as evidenced in
Table~\ref{tbl_cluster-dipoles} is probably the one principal
advantage conveyed by using a model with polarizable
ions. Classical models usually establish a molecular dipole
moment of the {\it monomer} to about 2--3~D at the outset of the
construction of the model.

Table~\ref{tbl_water-properties} lists some properties of liquid
water arising from our simulations, compared to previous GGA
calculations and experimental results. The static dielectric
constant is calculated using the method of Sharma~{\it et
  al.}\cite{Sharma07} and the self diffusion coefficient is
extracted as the linear slope of the mean square displacement. We
should note that both models give a reasonable prediction of the
dielectric constant, the genetic being rather better, which we
attribute to the anion polarizability.  The intuitive point
charge model overestimates $D_{\text{self}}$ which is consistent
with recent SCC-DFTB simulations,\cite{Maupin10} although these
latter overestimate by a great deal more, finding
$D_{\text{self}}=11.1\pm 0.4$ and~$6.5\pm 0.2\times 10^{-5}$
cm$^2$/s in two different modifications of the SCC-DFTB model,
which are very much greater than our $3.5\times 10^{-5}$
cm$^2$/s. This is very encouraging, firstly indicating that our
intuitive point charge model is better in this regard than the
SCC-DFTB model; more importantly by extending the theory to allow
anion polarizability results in a significant improvement in the
rendering of the self diffusion in the liquid. In contrast to the
GGA method,\cite{Sharma07} we reproduce $D_{\text{self}}$ quite
accurately without recourse to artificial elevation of the
temperature.

\section{Discussion}
\label{sec_discussion}

\subsection{General remarks on the new model}
\label{subsec_remarks}
The central result of this work is that it is possible to
construct TB models of water in a ``ground up'' approach; that
is, by fitting parameters to observed and calculated properties
of the monomer and dimer alone the model is found to be
sufficiently robust and transferable to be able to predict many
of the known properties of water. We believe that the rather
transparent construction that we have outlined in
section~\ref{sec_int} allows the models to provide some insight
into the very subtle properties of water, many of these being
manifestations of the hydrogen bond. In
section~\ref{subsec_H-bond} we made some observations on the
nature of the hydrogen bond, asserting that its principal two
components are an attractive force arising from the mutual
static polarizability of the monomers countered by a short ranged
repulsion which we model with the usual pair potential of TB
theory. Of secondary importance is the additional attraction due
to orbital overlap which is modeled by hamiltonian hopping matrix
elements as in all electronic structure localized orbitals
methods, including the density function description. We have not
addressed the question of van der Waals forces. While we expect
these to be weaker and to decay more rapidly with O--O distance
than the static polarizability, there is recent evidence from DFT
studies that their inclusion significantly improves upon the
local density approximation.\cite{Santra08,Wang10} One could
argue that by careful fitting to the CCSD(T) curve in
Fig.~\ref{fig_OO} we have ``renormalized'' our pair potential to
include some ``$1/r^6$-like'' attraction as would be done in a
classical potential. Conversely, as in the DFT, we could extend
our TB theory to include dynamic polarizability
explicitly.\cite{Stone96} Whether this can lead to an improved
description of the hydrogen bond must be left for future work. 

\subsection{The density of water and ice}
\label{subsec_density}
The most striking and anomalous feature of water is the fact that
ice is less dense than its liquid, leading to the possibility of
life on earth. Therefore we would very much like our models to
display this. Unfortunately as we have seen in
table~\ref{tbl_iceXI-density} our genetic dipole model gives an
ice density of 0.97~g/cm$^3$ at 0$^{\deg}$K. We find in an NPT
simulation at --3$^{\deg}$C that this decreases somewhat to
0.96~g/cm$^3$ but this is still greater than the average density
of the liquid in the NPT simulations at 27$^{\deg}$C, namely
0.926~g/cm$^3$. On the other hand this situation is nothing like
as bad as the GGA densities reported in the
literature.\cite{Hirsch04,Umemoto04,Wang10} These find the
density of ice-XI to be as much as 1~g/cm$^3$ while the liquid
density is as low as 0.8~g/cm$^3$ in the BLYP functional and
0.88~g/cm$^3$ in the PBE functional. This situation is rescued
using a functional including dynamic van der Waals forces but at
the expense of a poor rendering of the O--O~RDF.\cite{Wang10}
From the DFT point of view the question of the densities of ice
and liquid water is clearly not yet resolved. In the TB framework
it is most surprising and at the same time very encouraging that
the intuitive point charge model correctly predicts that ice will
float on water. This raises two questions. ({\it i\/}) Is the
inclusion of anion polarization an unnecessary and uncontrollable
complexity? In answer, we can certainly see from the above that
the genetic dipole model performs better than the point charge
model in respect of the RDF and properties of the monomer and
dimer. On the other hand we have not refined the point charge
model but have kept it at the ``intuitive'' level.  ({\it ii\/})
Could a point charge model be constructed that will adopt the
benefits of the dipole model and retain the correct difference
between the liquid and ice densities? This must be a subject of
further research but we suspect not. First we know from the
failings of the SCC-DFTB simulations,\cite{Maupin10} in respect
of both O--O RDF and diffusivity that the TB description can fail
at the point charge level, although our intuitive point charge
model corrects the worse shortcomings of the SCC-DFTB. Second,
and this is a point we wish to reemphasize, it is {\it only by
  including anion polarizability} that the trend of increasing
molecular dipole moment with cluster size into the limit of the
solid and liquid can be reproduced
(table~\ref{tbl_cluster-dipoles}). We believe that it is this one
element that elevates our TB model into direct competition with
the very much more (computer) time consuming DFT models.

We can use some of the results above to speculate upon the
question, why is ice less dense than water? The conventional
wisdom has it that hexagonal ice having four hydrogen bonds to
each molecule is less densely packed than the liquid in which, as
we see from the lowest panel in Fig.~\ref{fig_rdf}, each molecule
has on average between five and six neighbouring
molecules. However this cannot be the whole story because this is
true also in our genetic model and indeed published DFT
simulations in which ice is (wrongly) more dense than water. It
is important to note that, in contrast to the solid state, the
average O--O bond length in liquid water in our models depends on
the model parameters, but not strongly on the size of the
confining box, that is, on the volume. This is evident from the
first peaks of $g_{\text{OO}}(r)$ from the NVE and NPT
simulations in Fig.~\ref{fig_rdf}. Some light is thrown on the
matter by table~\ref{tbl_OO-lengths} which shows (average)
hydrogen bond lengths from our calculations using the genetic
dipole and intuitive point charge models. The point charge model
gives the bond length and hence density of ice-XI much more
accurately compared to experiment than the genetic dipole
model. Furthermore since the first peak of $g_{\text{OO}}(r)$
in the intuitive point charge model is close to the experiment
(Fig.~\ref{fig_rdf}) the consequence is that this model, perhaps
fortuitously, predicts the right ordering of the densities of ice
and water. This may well indicate that this will be correctly
rendered by any model that simultaneously predicts {\it both} the
density of ice {\it and} the position of the first peak of
$g_{\text{OO}}(r)$. However it is not obvious that such a model
can be easily found without specifically fitting these two
properties---this is of course possible for classical models.

\begin{table}
\begin{center}
\caption{\label{tbl_OO-lengths} O--O bond lengths in the
  structures studied here using our two models (in atomic units of
  length). We have used the cyclic hexamer as an example. In the
  case of ice-XI and liquid water these are averages.}
\begin{tabular}{lcccc}
\hline
 model & dimer & hexamer & ice-XI & liquid \\
\hline
genetic  & 5.51 & 5.00 & 5.13 & 5.52 \\
point    & 5.34 & 5.01 & 5.22 & 5.26 \\
\hline
\end{tabular}
\end{center}
\end{table}

\section{Conclusions}
\label{sec_conclusions}
We have demonstrated models for water at ``intuititive'' and
fitted levels in the point charge transfer and dipole charge
tight binding approximations. They have the benefit of quite
clear physical and chemical insight by virtue of the way they are
constructed, and at the same time possess quite remarkable
predictive power. This is evident from the ``ground up'' approach
that we have adopted in which only properties of the monomer (in
respect of O--H interactions) and the dimer (in the modeling of
the hydrogen bond) are fitted. Properties of clusters, ice and
liquid water are then predicted with good quantitative agreement
with experiment and published DFT calculations. Indeed in some
notable respects our models are superior to the DFT, in
particular in the cohesive energy curve of the dimer
(Fig.~\ref{fig_OO}) and in the density of ice-XI and the liquid
(table~\ref{tbl_iceXI-density}) which are particularly poorly
described by the GGA.

We expect our model to find a niche where nanosecond simulations
of some thousands of particles are called for while retaining a
proper self consistent quantum mechanical description of the
chemical bond. In particular, we believe that the model will
provide insight into the complexities of the hydrogen bond and in
the role of polar protic solvents in chemical reactions.

\begin{acknowledgments}
We are grateful to C.~Walsh and J.~Armstrong for contributions to
the early stages, and A.~Y.~Lozovoi to the later stages of this
work. We have enjoyed many stimulating discussions with
R.~M.~Lynden-Bell. We thank M.-V.~Fern{\'a}ndez-Serra for
providing an equilibrated sample of 128 water molecules; and
B.~Santra for the hexamer structure parameters. Financial support
was provided by the UK EPSRC under grant number EP/G012156/1.
\end{acknowledgments}

\appendix
\section{Generalized Madelung matrix}
\label{App_A}

The generalized Madelung matrix $\TBstrx$ is proportional to the
KKR structure constant matrix, $B$,\cite{Williams79,Andersen84}
\begin{equation*} 
\TBstrx_{\Rp\Lp\,\R\L}=\frac{4\pi}{\ls\lps}\>B_{\Rp\Lp\,\R\L}
\end{equation*} 
where
\begin{equation*} 
B_{\Rp\Lp\,\R\L}=4\pi\sum_{\Lpp}(-1)^{\l}
                   \lppms\>
                   C_{\Lp\L\Lpp}\>K_{\Lpp}(\R-\Rp)
\end{equation*} 
and
\begin{equation*} 
K_L({\bf r})=r^{-\l-1}\>Y_L({\bf r})
\end{equation*} 
is the solid Hankel function.
\begin{equation} 
\label{eq_Gaunt}
C_{\Lpp\Lp\L}
   =\int\!\!\!\!\int\!{\rm d}\Omega\>Y_{\Lpp}\,Y_{\Lp}\,Y_L
\end{equation} 
are the Gaunt integrals for real spherical
harmonics.\cite{Stone96}

\def\JPCM{J.~Phys.:~Condens.~Matter.}
\def\PRB{Phys.~Rev.~B}
\def\PRL{Phys.~Rev.~Lett.}

\end{document}